\documentstyle[aps,prb,epsfig,multicol]{revtex}

\def\ltsim{\mathop{\raise3pt\hbox{$<$}\llap{\lower3pt\hbox{$\sim$}}}}
\def\gtsim{\mathop{\raise3pt\hbox{$>$}\llap{\lower3pt\hbox{$\sim$}}}}
\newcommand{\chemie}[1]{$\rm #1$}

\newcommand{\journal}{\em}
\newcommand{\nature}[1] {{\journal Nature} {\bf #1}}

\begin{document}

\draft

\title{Dynamical properties of two doped, coupled Hubbard chains}

\author{R.M.\ Noack}

\address{Institut de Physique Th\'eorique, Universit\'e de Fribourg,
  CH-1700 Fribourg, Switzerland}

\author{M.G.\ Zacher, H.\ Endres, and W.\ Hanke } 

\address{ Institut f\"ur Theoretische Physik, Universit\"at
  W\"urzburg, \\ Am Hubland,  97074 W\"urzburg, Germany}

\maketitle
\begin{abstract}
Using quantum Monte Carlo (QMC) simulations combined
with Maximum Entropy analytic continuation
as well as analytical methods, we examine the one- and two-particle
dynamical properties of the Hubbard model on two coupled chains
at small doping. 
The behavior of the single-particle spectral weight 
$A({\bf k},\omega)$ as a function of hopping anisotropy $t_\perp/t$ at
intermediate interaction strength is dominated by the
transition from one-band behavior at large $t_\perp/t$ to two-band  
behavior at small $t_\perp/t$, although interaction 
effects such as band-narrowing, a shift of spectral weight to higher 
energies in the unoccupied antibonding band and reflected structures 
due to short-range antiferromagnetic correlations are also present.
A single-particle gap is resolved in the intermediate $t_\perp/t$
Luther-Emery phase using Density Matrix Renormalization Group
calculations.
The dynamical spin and charge susceptibilities
show features of the expected bonding-band 
Luttinger liquid behavior, as well as higher-energy features due to
local excitations between the chains at large $t_\perp/t$, 
and evolve towards the behavior of two uncoupled chains as $t_\perp/t$ is 
reduced.
For the one hole, large $t_\perp/t$ case, we make a detailed
comparison between the QMC data and an approximation based on local
rung states.
At isotropic coupling and somewhat larger doping, we
find that the dispersion of the single-particle bands is essentially
unrenormalized from that of the noninteracting system, and that
the spin and charge response functions
have features also seen in random-phase-approximation calculations.

\end{abstract}

\begin{multicols}{2}

\section{INTRODUCTION}
\label{INTRO}

In the last five years, there has been much experimental and
theoretical interest in systems of pairs of strongly correlated
spin--1/2 chains. \cite{dagrice}
Materials such as \chemie{SrCu_2O_3} \cite{aht94} \chemie{LaCuO_{2.5}} 
\cite{hit95}, and \chemie{Sr_{14}Cu_{24}O_{42-\delta}} \cite{uehara}
contain Cu--O planes which are periodically distorted leading to 
magnetically isolated two--chain Cu$_2$O$_3$ ladders with
approximately isotropic magnetic coupling.
These materials are spin--liquid insulators, i.e. they have
short--range antiferromagnetic order and a gap in the spin excitation
spectrum.
Recently, the Vanadium-based materials CaV$_2$O$_5$, MgV$_2$O$_5$ and
$\alpha'$--NaV$_2$O$_5$ have been proposed to contain V--O structures
similar to the Cu--O ladder structures. \cite{mila}
While CaV$_2$O$_5$ and MgV$_2$O$_5$ are spin liquid insulators, 
\cite{CaVO,MgVO} NaV$_2$O$_5$ is a one-dimensional antiferromagnet
that undergoes a spin-Peierls transition. \cite{NaVO}
The organic copper compound Cu$_2$(C$_5$H$_{12}$N$_2$)$_2$Cl$_4$ also
appears to be a two-chain ladder material with
antiferromagnetic spin-liquid properties.\cite{CuCl}
Many of the magnetic properties of these undoped, insulating materials
such as the presence and size of the spin gap, the magnetic
susceptibility, and the magnetic excitation spectrum, are
well--described by the Heisenberg model on an isolated ladder. 
\cite{barnes,etc,dagrice}
However, interladder coupling has been argued to be important for the
MgV$_2$O$_5$ material,\cite{MgVO} and a frustrating next-nearest-neighbor
interaction as well as a larger interchain than intrachain coupling for
Cu$_2$(C$_5$H$_{12}$N$_2$)$_2$Cl$_4$.\cite{CuCltheory} 

There have also been attempts to synthesize doped ladder materials.
Hiroi and Takano \cite{hit95} have succeeded in doping
the material \chemie{LaCuO_{2.5}} by substituting Sr for La.
While the doped material shows a finite spin gap and 
metallic behavior at low temperatures, there is no
transition to a superconducting state, which had been suggested
theoretically.\cite{drs92,rgs93,nws95,hpn95}
However, band structure calculations indicate that this material
probably has a substantial three-dimensional coupling between the
chains which would make an isolated ladder picture
invalid.\cite{normand,troyer}
The \chemie{Sr_{14}Cu_{24}O_{42-\delta}} material has been doped with
Ca \cite{mccarsieg} and Ca and La. \cite{bellgroup}
At low doping, the spin gap present in the insulating parent compound
persists and the resistivity becomes more metallic.
At high doping levels and high pressures (3 GPa), the composition
\chemie{Sr_{0.4}Ca_{13.6}Cu_{24}O_{41.84}} undergoes a superconducting
transition at $T_c = 12$K.\cite{uehara}
Anomalously, the transition temperature drops rapidly if the
pressure is further increased.
One issue that remains unclear is the extent to which the spin gap persists
in the high-pressure superconducting composition.
Mayaffre {\it et al.}\cite{jerome} have found low-lying spin
excitations in the Cu-O ladders in the normal state of the
superconducting composition \chemie{Sr_{2}Ca_{12}Cu_{24}O_{41}}
using $\phantom{a}^{63}$Cu-NMR measurements, perhaps indicating 
two-dimensional behavior in the high--pressure
superconducting phase.

Although the $\alpha'$--NaV$_2$O$_5$ material was originally thought to
consist of  charge-disproportionated
one-dimensional V-O chains, recent X-ray
diffraction measurements and band structure calculations\cite{gros}
suggest that it may also be a doped ladder system.
The material would then have one electron per V--O--V rung,
corresponding to a quarter-filled conduction band, and an interchain
hopping that is two to three times the intrachain hopping.
A quarter-filled two-chain ladder system with large interchain hopping
can form an effective one-dimensional antiferromagnet via the
interrung magnetic coupling of electrons in the rung bonding state.

There is growing evidence that a spin-liquid state like that
found in the ladder materials is closely connected to the physics of
the high-$T_c$ superconductors.
For example, the high--$T_c$ compounds show signs of a spin gap
above the N\'eel temperature and above the superconducting 
transition temperature at small doping.\cite{highTcgap}
The ``stripe phases'' of the two-dimensional CuO$_2$ planes
in the high--$T_c$ cuprates,\cite{stripes} which recently have
received considerable attention, are in some sense, ladder--like
structures.
Also, the copper--oxide--based ladder materials are clearly closely related
to high--$T_c$ materials, since the ladder-like arrangements of
Cu-atoms are produced by line defects in CuO$_2$ planes.

The success of the two-chain Heisenberg model calculations at
explaining the magnetic properties of the insulating materials has
lead to the consideration of the two--chain Hubbard and $t$--$J$
models as a starting point for the doped materials.
In fact, predictions of possible properties based on these models
predated the existence of doped materials.
The most interesting of these predictions was the possible existence
of a spin--gapped superconducting state with a $d_{x^2-y^2}$--like
order parameter in isotropic or near-isotropic materials at low
doping. \cite{drs92,rgs93,nws95,hpn95}
While the presence of superconductivity in the
\chemie{Sr_{0.4}Ca_{13.6}Cu_{24}O_{41.84}} material seems to bear out
this prediction, the extent to which isolated ladder models are
directly applicable to doped materials is not yet clear.
The assumption of isolated ladders is probably not
justified in the doped \chemie{LaCuO_{2.5}} material, and the absence
of a spin gap in \chemie{Sr_{0.4}Ca_{13.6}Cu_{24}O_{41.84}} does not
fit in with the picture obtained from the two--chain models.
Within the two--chain Hubbard model, the relative strength of the
superconducting correlations, which can at most follow a power law
decay in one dimension, is quite sensitive to the model parameters,
especially the doping level and the hopping anisotropy.\cite{enhanced}
Since these parameters also vary from material to material,
it is crucial to explore fully the dependence of a wide variety of
properties of isolated ladder models on them.
Here we will consider the behavior of the single and two--particle
dynamical response functions of the two--chain Hubbard ladder as a
function of doping and hopping anisotropy.

The Hamiltonian for a ladder consisting of two coupled chains of
length $L$ can be written
\begin{eqnarray}
H&=&-t\sum_{i,\lambda\sigma} (c^\dagger_{i,\lambda\sigma}
               c^{\phantom{\dagger}}_{i+1,\lambda\sigma} + 
               \textrm{h.c.})
    -t_\perp \sum_{i,\sigma} (c^\dagger_{i,1\sigma}
               c^{\phantom{\dagger}}_{i,2\sigma} + \textrm{h.c.})
\nonumber\\
  &&+U\sum_{i\lambda} n_{i,\lambda\uparrow}n_{i,\lambda\downarrow},
\label{Ketten}
\end{eqnarray}
where $\lambda=1,2$ numbers the chains and $i=1\ldots L$ the rungs of
the ladder, $t$ is the hopping matrix element within a chain, and
$t_\perp$ the hopping matrix element between the chains.

There has been much work published in the last few years on the
behavior of the one-band Hubbard model on ladder systems.
One important approach is a weak-coupling method in which the 
ground state properties of the system are investigated as a function
of $t_\perp$ and $U$ starting from the limit of weak $U$,
using renormalization group techniques (see, for example, 
Refs. \onlinecite{fpt92,sol79}). 
Particularly relevant for realistic two-chain model calculations is
the phase diagram of Balents and Fisher \cite{baf95}, which
is calculated in the limit $U\rightarrow 0+$, but for arbitrary values
of $t_\perp$.
The resulting phases can be classified by the number of gapless spin
and charge modes within a bosonization picture.
Since symmetric and antisymmetric ($k_\perp=0$ and $k_\perp=\pi$)
modes are possible between the chains, there are 9 possible phases,
labeled C$m$S$n$ in Ref.\ \onlinecite{baf95}, with $m=0,1,2$ the number of
gapless charge and $n=0,1,2$ the number of gapless spin modes.
For example, the C1S0 phase has one gapless charge mode and a spin gap.
This classification scheme has become widely used to describe the phases 
of one-dimensional coupled-chain systems in general since it provides a 
useful common notation.
Of the nine phases possible in this classification scheme, seven are
present in the weak-coupling phase diagram of Ref. \onlinecite{baf95}.

One phase of particular interest occurs for most of
the region of the phase diagram for which two bands cross the Fermi
energy, i.e. $t_\perp/t < 1 + \cos \pi \langle n\rangle$ for the
infinitely large noninteracting system.
The lightly to moderately doped isotropic system falls within this region.
This phase has one gapless charge mode and a finite spin gap (C1S0),
and is analogous to a one-dimensional Luther-Emery liquid.
A Luther-Emery liquid has power-law decaying 
charge-density-wave correlations (CDW) and singlet
superconducting correlations whose relative strength is governed by a
non-universal parameter.\cite{lue74,ima93,voi96}
With dominant superconducting correlations, the Luther-Emery liquid
is a one-dimensional analog of a superconductor, with
long-range superconducting order possible given three-dimensional
crossover behavior at low temperatures in the real materials.

Numerical studies of the two-chain Hubbard model using the
density-matrix renormalization group (DMRG) technique
\cite{nws95,nws94} show that for a realistic value of the
Coulomb interaction, $U=8t$, the doped system does indeed exhibit a spin gap
and metallic behavior, i.e.\ a vanishing charge gap, in approximately
the same region of the $t_\perp$-$\langle n \rangle$ phase diagram as
the C1S0 phase in the weak coupling calculation of Balents and
Fisher.\cite{baf95}
In addition, there is a direct correlation between the strength of the
pair-pair correlation function and the size of the spin
gap.\cite{nws95}
The pair order parameter possesses a modified $d_{x^2-y^2}$ symmetry
and the pair correlation decays as a power law.
These results are consistent with those for a $t$-$J$ model in the
small $J/t$ region where the strong $U$ mapping of the Hubbard model
applies.\cite{hpn95}
However, the pairing correlation function does not seem to decay
slowly enough for isotropic coupling between the chains
in order for the superconducting correlations to be the dominant
correlations.

Dynamical response functions are important both because they can be 
directly probed by experimental techniques and because they provide a 
detailed picture of the excitation spectra at different energy scales, 
which can often be understood qualitatively or quantitatively through 
various theoretical pictures or approximations.
In particular, the single particle spectral weight function 
$A({\bf k},\omega)$ can be measured directly in photoemission or inverse 
photoemission experiments, and the two--particle response can be 
probed using, for example, inelastic neutron--scattering experiments.

Previously, work on the dynamical excitations of the doped two-chain
system has concentrated mostly on the $t$--$J$ model.
In Ref.\ \onlinecite{ttr96}, the photoemission spectra and
dynamical spin and charge excitations in the half--filled and lightly
doped $t$--$J$ ladder were studied using exact diagonalization.
The starting point of large interchain $J$ was used to understand many
features of the spectra and the isotropic case was studied extensively.
In Ref.\ \onlinecite{haasdagotto}, the single-particle spectral weight
of the lightly doped isotropic $t$--$J$ ladder was studied using exact
diagonalization, with an emphasis on magnetic features present in the
single--particle spectrum, and the effect of an interchain Coulomb
repulsion. 
Recently, Riera, Poilblanc and Dagotto have considered the the effect
of the anisotropy on the photoemission spectrum in both the Hubbard
and $t$--$J$ models using exact diagonalization.\cite{anisodyn}
Endres {\it et al.} \cite{enh96} studied both the single particle and
two--particle dynamical correlation functions of the half--filled
Hubbard model using quantum Monte Carlo (QMC) techniques and a Maximum
Entropy analytic continuation (MaxEnt), emphasizing the behavior as a
function of the hopping anisotropy, and making a comparison to various
analytic calculations.
The purpose of the present work is to extend the QMC/MaxEnt
calculations to the doped case.
Some of the work is contained in unpublished form in
Ref.\ \onlinecite{endresphd}, and a few of the results were published in
preliminary form as part of a conference proceedings.\cite{endresconf}

Here we will calculate both the single--particle spectral weight
function, $A({\bf k},\omega)$, and the dynamical two--particle spin
and charge susceptibilities, $\chi_s({\bf q},\omega)$ and 
$\chi_c({\bf q},\omega)$.
It is important to examine both types of correlation functions in
order to make contact with the weak--coupling phase diagram in which,
in general, well--defined quasiparticles do not exist, but the phases
can be understood in terms of bosonic spin and charge excitations.
The single--particle spectral weight makes contact with the original
band structure and its evolution with interaction, and the
two--particle dynamical correlations contain information on the nature
of the low-lying bosonic magnetic and charge excitations of the system.
The single--particle spectral weight function is
defined in the Lehmann representation as
\begin{eqnarray}
A({\bf k},\omega) &=& {1\over Z}
\sum_{l,l^\prime} e^{-\beta E_l}(1+e^{-\beta\omega})
\vert \langle l\vert c_{{\bf k},\uparrow}\vert l^\prime \rangle \vert^2
\nonumber\\
&& \phantom{{1\over Z}\sum_{l,l^\prime}}
\cdot\delta(\omega-(E_{l^\prime}-E_l)),
\label{Akomegadef}
\end{eqnarray}
where $Z$ is the partition function, $\vert l>$ is
the exact many-body eigenstate with the energy $E_l$,
$\beta$ is the inverse temperature $1/k_B T$ and
$c_{{\bf k},\sigma}=\sum_{i,\lambda}
c_{i,\lambda\sigma} e^{i{\bf k}\cdot{\bf R}_{i,\lambda}}/\sqrt{2L}$.
Throughout this paper, we will use the notation ${\bf k}=(k_\perp,k)$
for momenta, where the transverse component, $k_\perp$, can take on
the values 0 and $\pi$, and $k$ denotes the component parallel to the
chains.
We use the grand-canonical Quantum Monte Carlo (QMC) algorithm to
determine the expectation values of the correlation functions in
imaginary time, and invert the spectral theorem
\begin{equation}
  G({\bf k},\tau)\equiv \langle c^{\phantom{\dagger}}_{{\bf k},\uparrow}(\tau)
  c^\dagger_{{\bf k},\uparrow}(0) \rangle
  = \int_{-\infty}^{\infty} 
  {e^{-\tau\omega}\over 1+e^{-\beta\omega}}
  A({\bf k},\omega) d\omega
\label{eqGtaudef}
\end{equation}
using the Maximum Entropy analytic continuation method (MaxEnt).\cite{maxent}
In order to achieve high resolution, we use a likelihood
function which takes the error--covariance matrix of the QMC data and
its statistical inaccuracy consistently into
account\cite{preussmaxent}, and incorporate various moments of
the spectral weight.\cite{whi91}
While the fermion sign problem is present, we have chosen system
parameters, i.e. $U/t$, filling $\langle n \rangle$, and lattice size
$L$ so that low temperatures can be reached with sufficient accuracy.
The results presented here are based on QMC data with good statistics,
i.e. averages over $10^5$ updates of all the Hubbard--Stratonovich
variables result in $G({\bf k}, \tau)$'s with absolute errors less
than or of the order of $5 \times 10^{-4}$.

The two-particle correlation functions are defined similarly in the
Lehmann representation, 
\begin{eqnarray}
\chi_{s,c}({\bf q},\omega)&=&
   {i\over Z}\sum_{l,l^\prime}e^{-\beta E_l}
   (1-e^{-\beta\omega})
   \vert\langle l\vert O_{s,c}({\bf q})\vert l^\prime\rangle\vert^2
\nonumber\\
   &&\phantom{{i\pi\over Z}\sum_{l,l^\prime}}
   \cdot\delta(\omega-(E_{l^\prime}-E_l)),
\label{eqtwocorrdef}
\end{eqnarray}
with $O_{s}({\bf q})=\sum_{\bf p}
(c^\dagger_{{\bf p}+{\bf q},\uparrow}c^{\phantom{\dagger}}_{{\bf p},\uparrow}
-c^\dagger_{{\bf p}+{\bf q},\downarrow}
c^{\phantom{\dagger}}_{{\bf p},\downarrow})$ 
and
$O_{c}({\bf q})=\sum_{{\bf p},\sigma} c^\dagger_{{\bf p}
+{\bf q},\sigma}c^{\phantom{\dagger}}_{{\bf p},\sigma}$. 
The calculation in imaginary time and the analytic continuation are
analogous to the calculations for $A({\bf k},\omega)$.

Although the QMC results are relatively high resolution, some
low-energy properties such as single--particle and spin gaps are at an
energy scale of the order of $J=4t^2/U$, and are not resolvable with
QMC/MaxEnt.
In order to complement the QMC calculations, we will also present some Density
Matrix Renormalization Group \cite{whitedmrg} (DMRG) results for
low-energy gaps and static correlation functions. \cite{nws95}
The DMRG can provide very high resolution results for such quantities,
but cannot yet provide information on the full dynamical properties.

This paper is organized as follows.
In Sec.\ \ref{tpevol}, we will examine the general evolution of
both $A({\bf k},\omega)$ and $\chi_{s,c}({\bf q},\omega)$ as a
function of $t_\perp/t$ at $U/t=4$ and moderate doping, 
$\langle n \rangle =0.94$, i.e. two holes on a $2\times 16$ lattice.
We emphasize making contact with the band transition as a function of
$t_\perp/t$ as well as the weak--coupling phase diagram.
DMRG calculations of the single--particle gap and the static spin
structure factor will be used to resolve low-energy features.
In Sec.\ \ref{onehole}, the case of one hole ($\langle n
\rangle=0.9875$ on a $2\times 16$ lattice) will be treated at $U/t=8$
and large $t_\perp/t=2.0$
Here a direct comparison will be made with a local rung approximation
(LRA), a strong-coupling expansion based on the states of an isolated
rung \cite{enh96,barnes,ttr96}.
We find very good agreement for the two--particle as well as
the single--particle dynamical correlation functions.
Finally, we will treat the isotropic system, $t_\perp/t=1.0$ at
a higher doping $\langle n \rangle = 0.875$ and $U/t=4.0$ in 
Sec.\ \ref{isodope}.
This case is interesting since the system is somewhat further from
half--filling, and the antiferromagnetic correlations are weaker.
We will compare the two--particle dynamical correlations with Random
Phase Approximation (RPA) calculations, and the single--particle
spectral weight function with the noninteracting bands and with slave
boson mean--field calculations.
We find that the dispersion of the bands is essentially unrenormalized
from that of the noninteracting system and that the RPA calculations
lead to a qualitative understanding of some features of 
$\chi_s({\bf q},\omega)$ and good quantitative agreement with most of
the features $\chi_c({\bf q},\omega)$.
The points in the zero-temperature $\langle n \rangle$--$t_\perp/t$ phase
diagram at which we make calculations are shown in Fig.\ \ref{U0phase} as 
points in the $U=0$ phase diagram.

\narrowtext

\section{Evolution with \lowercase{$t_\perp$/t}}
\label{tpevol} 

In this section, we discuss the evolution of the single-particle
spectral weight $A({\bf k},\omega)$, and the dynamic spin and charge
susceptibilities $\chi_s({\bf q},\omega)$ and $\chi_c({\bf q},\omega)$,  with
decreasing $t_\perp/t$ at low doping, $\langle n \rangle = 0.94$.
We choose intermediate coupling, $U/t=4$, and show results on a
$2\times 16$ lattice for which $\langle n \rangle = 0.94$
corresponds to a doping of two holes from half-filling.
The inverse temperature, $1/T \equiv \beta = 8t$ is low enough so that
the system is effectively in the ground state for this system size, 
to within the resolution of the QMC calculations and the MaxEnt 
analytic continuation.

As we will see below, the behavior of $A({\bf k},\omega)$ is dominated by
the band transition as a function of $t_\perp/t$.
The dispersion of the bands is only moderately renormalized by
band-narrowing interaction effects, and the one-band to two-band
transition takes place at $t_\perp/t=1.6$, a significantly smaller
value than the noninteracting ($L\rightarrow\infty$) value of
$t_\perp/t=1.98$.
There are also additional features present due to the
interactions.

We display $A({\bf k},\omega)$ in Figs.\ \ref{Akwtp2p2}--\ref{Akwtpp2} for
$t_\perp/t=2.2,1.6,1.0$ and 0.2.
The band transition can be clearly seen.
For $t_\perp=2.2$, only the $k_\perp=0$ bonding band lies below the
Fermi level, and the dispersion of the clearly defined peak follows
that of the $U=0$ band (shown as a dotted line).
The Fermi wave vector within this band is given by 
$k_F^b = \pi\langle n \rangle = 15\pi/16$.
In the antibonding ($k_\perp=\pi$) channel, the spectral weight is in
the unoccupied $\omega >0$ portion of the spectrum and lies mostly
above the noninteracting band.
The total spectral weight is small, and while the peaks have
approximately the dispersion of the noninteracting band, they are
shifted to higher energy and are quite broad.
In addition, there are some reflected structures in this channel near
$k=0$ probably due to short-range antiferromagnetic order.

At $t_\perp/t=1.6$ (Fig.\ \ref{Akwtp1p6}) the interacting system is
just below the band transition.
The bonding band is almost completely occupied and the antibonding
band is almost completely unoccupied, leading to $k_F^a \approx 0$ and
$k_F^b \approx 15 \pi/16$.
There is band flattening in both bands near the Fermi points,
leading to a higher density of states in these regions.
This high density of states near the Fermi level separated by a
wave vector ${\bf Q}=(\pi,\pi)$ was argued in Ref.\ \onlinecite{enhanced} to
strongly enhance $d_{x^2-y^2}$--like pairing correlations in this
regime.
According to the weak coupling phase diagram of 
Ref.\ \onlinecite{baf95} and the numerical DMRG calculations of 
Ref.\ \onlinecite{nws95}, this phase is a C1S0 phase, 
a Luther-Emery liquid with a spin gap.
Structures resembling the reflected bands present at half--filling are
also seen, probably reflecting the short-range antiferromagnetic order
present close to half-filling.

As $t_\perp/t$ is further reduced, as seen in Fig.\ \ref{Akwtp1}, the
occupation of the antibonding band increases and that of the bonding band
decreases.
Below the Fermi level ($\omega < 0$), the dispersion of the bands
follows the noninteracting dispersion well.
There is some flattening of the bands at the Fermi points and
reflected structures can also be seen.

At $t_\perp/t=0.2$, Fig.\ \ref{Akwtpp2}, the bonding and antibonding
bands are almost equally occupied, and the distribution of spectral
weight is very similar in the two bands.
There is a lightly weighted broad reflected structure in the
$\omega > 0$ portion of the spectral weight, and a broad incoherent
background at low energies for $\omega < 0$ in both $k_\perp$ channels.
At these parameters, $2 k_F^b \approx \pi$, a condition which leads to
a C1S2 phase (i.e. gapless spin and charge excitations) in the
weak-coupling phase diagram of Ref.\ \onlinecite{baf95}.

It is interesting to determine to what extent the characteristics of
the low-energy behavior are reflected in $A({\bf k},\omega)$.
Since some of the phases predicted as a function of $t_\perp$ are
well-understood one-dimensional phases such as a Luttinger liquid
(C1S1) and Luther-Emery liquid (C1S0) phases, characteristics of these
phases should be present.
In principle, a Luttinger liquid phase should lead to two separate
anomalous peaks in $A({\bf k},\omega)$ for $k$ near $k_F$, corresponding to
the separated spin and charge modes.\cite{luttdyn}
The Luther-Emery phase should lead to a single-particle gap of the
order of the spin gap, as well as an anomalously diverging peak
corresponding to the charge mode.\cite{voi96}
However, resolving the spin-charge separated peaks with numerical
methods has proven to be rather difficult, even for the numerically
easier case of the one-dimensional Hubbard model.\cite{spinchargesep}
There are no separately resolvable spin and charge peaks or resolvable
single-particle gaps in Figs.\ \ref{Akwtp2p2}--\ref{Akwtpp2}.
The DMRG, however, can be used to resolve gaps at quite small energy
scales.

In Fig.\ \ref{quasigap}, we display the finite-size scaling of the
single-particle gap 
\begin{equation}
  \Delta_1 = \mu^+ - \mu^- = E_0(N+1) + E_0(N-1) - 2 E_0(N)
\end{equation}
calculated on $2\times 8$ to $2\times 32$ lattices
with open boundary conditions.
Here $\mu^+ - \mu^-$ is the jump in the chemical potential at the
Fermi surface, and $E_0(N)$ is the ground state energy with minimum
$|Sz|$ and $N$ electrons at a particular system size, and the overall
density $n = N/L$ is kept fixed for different system sizes. 
The lines are least-squares fits to quadratic polynomials in $1/L$.
We have chosen somewhat larger doping, $\langle n \rangle = 0.875$
than the QMC calculations in order to have more system sizes available
for the finite-size scaling, and $U/t=8$.
The values $t_\perp/t=1.3,1.6$ have been chosen to be just below and
just above the two-band to one-band transition, as determined from
DMRG calculations of the spin gap and the pair correlations. \cite{nws95}
As can be clearly seen, there is a finite value of $\Delta_1$ at
$L\rightarrow\infty$ for $t_\perp/t=1.3$, in agreement with the
prediction for a Luther-Emery system in Ref.\ \onlinecite{voi96}.
For $t_\perp/t=1.6$, $\Delta_1$ scales to zero as
$L\rightarrow\infty$, as is predicted for the Luttinger liquid (C1S1)
phase which should be present at these parameters.

The behavior of the spin and charge modes can more directly examined
through the dynamical spin and charge response 
$\chi_s({\bf q},\omega)$ and $\chi_c({\bf q},\omega)$.
We exhibit QMC/MaxEnt calculations of these quantities in 
Figs.\ \ref{Chitp2p2}--\ref{Chitpp2} for the same parameters as in
Figs.\ \ref{Akwtp2p2}--\ref{Akwtpp2}.

For $t_{\perp}/t=2.2$, Fig.\ \ref{Chitp2p2}, the system is predicted 
to be a Luttinger liquid (C1S1 phase) in the bonding band since the 
antibonding band is completely unoccupied.
Therefore, one would expect to see gapless spin and charge 
excitations in the $q_{\perp}=0$ channel at 
$q=2k_{F}^b=2\pi\langle n \rangle$.
The presence of such low-lying excitations at $q=\pi/8$ (where the
wave vector has been wrapped back to the first Brillouin zone) can be
seen in Fig.\ \ref{Chitp2p2} (a) and (c). 
The dispersion is consistent with a linear increase, as would be 
expected in a Luttinger liquid, although the the resolution is not 
sufficient to resolve different spin and charge velocities.
The presence of a spin correlation at ${\bf q}=(0,2k_F^b=\pi/8)$
can also be seen as a small peak in the $q_\perp=0$ branch of the spin
structure factor, $S({\bf q})$, plotted in Fig.\ \ref{Sqn94}.
Here we have calculated $S({\bf q})$ at $T=0$ on a $2\times 32$ lattice 
using the DMRG.
Note that the much more heavily weighted $q_\perp=\pi$ branch of 
$S({\bf q})$ is flat, indicating very strong antiferromagnetic
correlations across a rung of the ladder, but very short--range
correlations along the legs.

The majority of the spectral weight, however, is associated with
high-energy features in the $q_\perp=\pi$ channel for both 
$\chi_s({\bf q},\omega)$ and $\chi_c({\bf q},\omega)$.
For $\chi_s({\bf q},\omega)$, Fig.\ \ref{Chitp2p2}(b), there is a 
dispersive, cosine-like band with larger spectral weight near $q=0$ 
and a minimum near $q=\pi$.
This dispersive band is also seen in the half-filled system\cite{enh96} 
and can be understood in terms of a local rung approximation in which 
a triplet excited state on a rung is delocalized in a cosine band of 
width $J=4t^2/U$.
This picture, valid for large $t_{\perp}/t$ at half-filling, provides 
a good description of the $q_{\perp}=\pi$ spin excitations because of 
the proximity of the system to half-filling and due to lack of 
low-lying spin excitations in the antibonding channel.
The size of the $q_{\perp}=\pi$ spin gap is determined by the band 
separation $t_{\perp}-t_{\perp c}$ where $t_{\perp c}/t \approx 1.6$ 
for these parameters.

%

The $q_{\perp}=\pi$ charge susceptibility, Fig.\ \ref{Chitp2p2}(d), is 
also very similar to that of the half-filled system.
There is a strongly weighted peak at $q=0$ at $\omega/t \approx 7$
which can be understood in terms of a local charge excitation on a
rung.
As $q$ is increased, this charge excitation becomes a broad band with
decreasing spectral weight with increasing $q$.
The peak in intensity has a dispersion which increases with increasing
$q$.
There is, however, a small amount of spectral weight spread over a
region with a lower bound which decreases with increasing $q$.
This broadening of the region of charge excitations can be understood
principally in terms of particle-hole excitations between the single-particle
bands, although vertex effects do also play a role. 
A more detailed description of these effects for the half--filled case is 
given in Ref.\ \onlinecite{enh96}.

At $t_\perp/t=1.6$, the system is just inside the doped spin-liquid
(C1S0) phase, and a spin gap is present.
As can be seen in Fig.\ \ref{Chitp1p6} (a), the $q_\perp=0$ branch of 
$\chi_s({\bf q},\omega)$ has gained a little in spectral weight
relative to $t_\perp/t=2.2$ and the band of charge excitations has
become narrower in $\omega$.
While there is some spectral weight
near $\omega=0$ and $q=\pi/8$, as was seen for $t_\perp/t=2.2$, here
one would expect a gap in the spin excitations. 
The finite resolution here, however, precludes resolving a gap of the
order of $0.1t$ or less.
In the $q_\perp=\pi$ branch, Fig.\ \ref{Chitp1p6} (b), the cosine-like
dispersive band is still present and has the same bandwidth, $2t$, as for
$t_\perp/t=2.2$, but is shifted downwards in $\omega$ so that the
$q=\pi$ peak is close to $\omega=0$.
This shift of the band is expected since the gap in the $q=\pi$
excitations in the large $t_\perp$ regime is determined by the band
separation, and should go to zero at the band transition.
The minimal spin gap, of order $0.1t$, should occur at 
${\bf q}=(\pi,k_F^b + k_F^a=15\pi/16)$, but
the limited resolution in ${\bf q}$ and in $\omega$ does not allow us
to unambiguously resolve the position of of the minimum, or the size
of the gap.
The spectral weight at $q=\pi$ is larger than for
$t_\perp/t=2.2$, Fig.\ \ref{Chitp2p2} (b).



The behavior of the spin structure factor $S({\bf q})$, shown in 
Fig.\ \ref{Sqn94}, reflects the qualitative change in the spin
correlations due to the band transition.
For $t_\perp/t=1.6$, the peak at ${\bf q}=(0,\pi/8)$ present for
$t_\perp/t=2.2$ which was due to the $2 k_F^b$ correlations in the
C1S1 phase has disappeared, as one would expect.
There is a larger amplitude in the $q_\perp=0$ correlations, reflecting
stronger intrachain antiferromagnetic correlations, and a relatively
broad peak at $q=k_F^b + k_F^a = 15 \pi/16$ has developed.
This peak reflects short-range incommensurate spin fluctuations
present in the C1S0 phase. \cite{nws94}

The dynamic charge susceptibility, Fig.\ \ref{Chitp1p6} (c) and (d),
has only minimal qualitative changes from the $t_\perp/t=2.2$ case.
In the $q_\perp=0$ channel, $\chi_c({\bf q},\omega)$ has a minimum in
the dispersive band near $q=\pi/8$, with a dispersive band which is
somewhat broader and more heavily weighted than for $t_\perp/t=2.2$.
In the $q_\perp=\pi$ branch, the structures are similar to the
$t_\perp/t=2.2$ case, but with somewhat broader peaks, and, as
expected, are also shifted to lower $\omega$ due to the smaller separation
of the bonding and antibonding bands.
Note that selection rules forbid there being any spectral weight at 
${\bf q} =(0,0)$ for both $\chi_c({\bf q},\omega)$ and 
$\chi_s({\bf q},\omega)$.

As $t_\perp/t$ is further lowered to 1.0, Fig.\ \ref{Chitp1} and 0.2,
Fig.\ \ref{Chitpp2}, $\chi_s({\bf q},\omega)$ and 
$\chi_c({\bf q},\omega)$ evolve towards a form expected for two
uncoupled chains.
One can see that as $t_\perp$ is lowered, the $q_\perp=0$ and
$q_\perp=\pi$ branches of both the spin and charge susceptibilities
become similar to one another, until at $t_\perp/t=0.2$, the
branches are almost identical to within our resolution.
At $t_\perp/t=1.0$, one sees that the dispersion of the peaks in
$\chi_s({\bf q},\omega)$ has minima near $\omega=0$ at ${\bf q} = (0,0)$
and near ${\bf q} = (\pi,\pi)$, with heavy spectral weight at 
${\bf q} = (\pi,\pi)$.
The isotropic system should be in a spin-gapped C1S0 phase.
The existence of a spin gap has been confirmed numerically with DMRG
calculations \cite{nws95} for $U/t=8$ and is of the order of $0.08t$ at
$\langle n \rangle=0.9375$.
Again, this spin gap is smaller than the resolution available in the
QMC/MaxEnt calculations.


The charge susceptibility, Fig.\ \ref{Chitp1} (c) and (d), has a
broad, heavily weighted dispersive band with a minimum at $q=0$ and
$\omega \approx 0$ and a maximum near $q=\pi$ in the $q_\perp=0$ branch.
In the $q_\perp=\pi$ branch, there is also a dispersive band with a
broad peak whose position increases monotonically with increasing $q$.
However, the minimum position at $q=0$ is at $\omega \approx 3t$.
In addition, there are more lightly weighted dispersive structures at
smaller $\omega$ that reach $\omega=0$ at a number of finite wave
vectors in both $q_\perp$ branches.
The overall spectral weight distribution and the position of these
minima can be understood by comparing with Random Phase Approximation
(RPA) calculations of $\chi_c({\bf q},\omega)$.
Such a comparison will be carried out in detail for a larger doping,
$\langle n \rangle = 0.875$ in Sec.\ \ref{isodope}, where we have better
resolution data on a larger lattice.

%

At $t_\perp/t=0.2$, weak-coupling theory \cite{baf95} predicts that
the system is close to the line at which the bonding band is
half-filled at $U=0$, and Umklapp processes in the bonding band lead
to a C1S2 phase, i.e. a vanishing of the spin gap.
DMRG calculations have shown the spin gap vanishes along this line
even at large interaction, $U/t=8$.
Figs.\ \ref{Chitpp2}(a) and (b) are consistent with two vanishing spin
modes, to within the resolution of the QMC.
Both the $q_\perp=0$ and the $q_\perp=\pi$ branches have a sine-like
dispersion with minima at $q=0$ and near $q=\pi$.
The spectral weight distribution is similar in the two bands, except
for somewhat larger weight near ${\bf q} =(\pi,\pi)$, due to the
remaining antiferromagnetic correlations, and zero weight at 
${\bf q}=0$ due to selection rules.
The dispersion and weight distribution in each channel are similar to
that of the 1D Hubbard model. \cite{pml94}
The charge and spin velocities of the 1D Hubbard model, obtained from
Ref.\ \onlinecite{schulz}, are indicated in Fig.\ \ref{Chitpp2} as
lines at low $q$.
As can be seen, the low-$q$ dispersion agrees well with this 1D value for
both $\chi_s({\bf q},\omega)$ and $\chi_c({\bf q},\omega)$ in both
$q_\perp$ channels.

\section{One hole, large interchain coupling}
\label{onehole}

For small doping and $t_\perp\gg t$, it is possible to
approximate the properties of the two coupled Hubbard chains by
starting from a local rung-singlet picture. 
For vanishing doping, the single-particle spectral weight can be
calculated by considering transitions between a state with one
hole and a state with two holes.
A one-hole state with parallel momentum $k$ is given in the LRA
\cite{enh96} as
$\vert\psi_1(k)\rangle =\sum_{\ell=1}^L e^{ik\ell}\vert\ell\rangle/\sqrt{L}$.
The expectation value of the energy for this state in first
order perturbation theory is
\begin{equation}
E_1(k)= \langle \psi_1(k)\vert H \vert\psi_1(k) \rangle =
(L-1)E_a-t_\perp+2tA\cos k,
\end{equation}
with $A=(1+E_b/2t_\perp)^2/(2+E_b^2/2t_\perp^2)$. 
The lowest energy state thus occurs for $k=\pi$ with the wave vector
\begin{equation}
\vert\psi_0^{\rm 1 hole}\rangle = 
{1\over\sqrt{L}} \sum_{\ell=1}^L (-1)^\ell \vert S_1\rangle\vert
S_2\rangle\ldots\vert\downarrow_\ell\rangle\ldots\vert S_L\rangle,
\end{equation}
and energy $E_1^{\rm 1 hole}=(L-1)E_a-t_\perp-2tA$.
The wave function for this state is a product of rung eigenstates, all
of which have a momentum $k_\perp=0$ between the chains.
Therefore, the total perpendicular momentum $K_\perp$ is zero.
As discussed in Ref.\ \onlinecite{enh96} (see Fig.\ 7), it is energetically
most favorable to remove a particle from a two--particle rung state
$\vert S_{\ell}\rangle$ in the $k_\perp=0$ channel.
The removal of a second particle from the one--hole ground state
will predominantly take place on a rung with two particles and have 
$k_\perp=0$.
(The removal of the particle from $\vert\downarrow_\ell\rangle$ will
also be a $k_\perp=0$ process, since the state with no particle has 
$k_\perp=0$.)
Therefore, we will consider only processes in the
$k_\perp = 0$ channel here.
The LRA state for two holes can then be written as
\begin{equation}
\vert\psi^{\rm 2 hole}(k)\rangle = {1\over B_N} c_{k,k_\perp=0,\downarrow}
\vert\psi_0^{\rm 1 hole}\rangle,
\end{equation}
where $B_N$ is a normalization constant.
Without loss of generality, we can choose the direction of the spin so
that the $z$ component of the total spin vanishes.
After some calculation, one finds that the dispersion in the 
$k_\perp=0$ channel for $\omega<0$ is given by
\begin{eqnarray}
\omega^{\rm 1 hole}(k)=-2tA^{\rm 1 hole}(1+\cos k),
\label{LRA_dot_dis}
\end{eqnarray}
with $A^{\rm 1 hole}=(1+E_b/2t_\perp)^2/(2+E_b^2/2t_\perp^2)$. 
The factor $A^{\rm 1 hole}$, which renormalizes the band, is identical
to that obtained in the half-filled case (i.e.\ a transition from a
half-filled LRA state to a one-hole LRA state).
In other words, to first order in $t$, the effective bandwidth does not change
with doping near half-filling.

In Fig.\ \ref{xyz}, we compare the QMC and LRA results on a $2\times 16$
system with $U/t=8$ and $t_\perp/t=2.0$ for $k_\perp=0$, part (a), and
$k_\perp=\pi$, part (b).
The average particle density, $\langle n\rangle=0.96875$, corresponds
to a doping of one hole on a $2 \times 16$ lattice.
The inverse temperature, $\beta t=8$, is the maximum that could be
reached due to the fermion sign problem, but is nevertheless low
enough so that the $2 \times 16$ system is effectively in the ground
state, to within the resolution of the MaxEnt data.

The solid line in the $k_\perp=0$ channel shows the results from the
LRA calculation.
As in the half-filled case, the dispersion is in very good agreement
with the dispersion of the main peak in the QMC data,
even at $t_\perp/t=2.0$.
Note that the $U=0$ band, shown as a dotted line, has a larger width
and does not fit the peaks as well.
There is a broad band between 
$\omega\approx 4t$ and $\omega\approx 8t$ in the $k_\perp=\pi$-channel
whose bandwidth is the same as that of the band in the $k_\perp=0$
channel to within error bars.
This band is displaced upwards significantly from the $U=0$ band,
indicating substantial deviation from the noninteracting band structure.
Because the resolution of the MaxEnt method becomes less accurate for larger 
$\omega$ values, the dispersion for $k_\perp=\pi$ is not as easily
discerned as for the corresponding band in the $k_\perp=0$ channel.
While most of the spectral weight is contained in these dispersive
bands, there are indications of a broad incoherent background, similar
to those found at half-filling, in both $k_\perp$ channels.

In Fig. \ref{CS_r096875} we show the charge and spin susceptibilities
for the same parameters as above.
Almost all of the spectral weight in the spin response function is
contained in the $q_\perp=\pi$ channel.
One can understand this result by considering the Lehmann
representation of $\chi_s({\bf q},\omega)$ in the LRA picture.
Since for $q_\perp=0$ the operator $O_s({\bf q})$ commutes with the
LRA Hamiltonian $H_0$, the matrix elements 
$\langle l\vert O_s({{\bf q}})\vert l^\prime\rangle$ are diagonal in 
$l$ and $l^\prime$, the susceptibility
$\chi_s((q_\perp=0,q),\omega)$ is therefore exactly zero when $H_I$ is
neglected (i.e. for $t_\perp\gg t$).
In the $q_\perp=\pi$ channel, most of the spectral weight lies in a
band with a cosine-like dispersion (Fig.\ \ref{CS_r096875}(b))
between $\omega\approx 1.9t$ and $\omega\approx 0.8t$.
A band with identical form to within the error bars is found in the
half-filled case \cite{enh96}, and can also be identified with
magnon excitations with the LRA.
The spin gap in the $q_\perp=\pi$ channel is the same as
the half-filled value, $\Delta E_S^{\rm half-filling}/t=0.81\pm0.02$ to 
within the resolution of the QMC/MaxEnt data.
The weak-coupling phase diagram\cite{baf95}
predicts a Luttinger-liquid-like phase with gapless charge and spin
excitations (C1S1) in this parameter regime, i.e.\ no spin gap.
There is some indication of weakly weighted low-lying spin
excitations in the $q_\perp=0$ channel of the spin susceptibility
shown in Fig. \ref{CS_r096875}(a), supporting the
weak-coupling phase diagram even for the relatively strong coupling
$U/t=8$. 
Note that this feature is present with a larger weight for
$\langle n \rangle = 0.94$ and $U/t=4$, as we have seen in 
Fig.\ \ref{Chitp2p2}(a). 



In Figs.\ \ref {CS_r096875}(c) and (d), we exhibit the charge
susceptibility $\chi_c({\bf q},\omega)$ for the same parameters.
For momentum $q_\perp=\pi$, shown in Fig. \ref {CS_r096875}(c), there
is a broad, $q$-independent excitation at $\omega\approx 9t$.
This excitation can be understood in the LRA picture in terms of
charge excitations on one rung which are also present in the
half-filled case.
Due to the fermion sign problem, which becomes worse with doping,
it is not possible to obtain as high a resolution as for the
half-filled system.
The excitation spectrum for $q_\perp=0$ is, however, quite interesting
in that we find a well-defined $\omega\rightarrow 0$-Mode for
$q\rightarrow 0$ with velocity $v_c/t=0.59\pm 0.29$, even though there
should be no $q_\perp=0$ spectral weight within the LRA picture
[$O_c(q_\perp=0,q)$ commutes with the unperturbed Hamiltonian
$H_0$, analogously to $O_s(q_\perp=0,q)$ for the spin excitations].
The presence of low-lying charge excitations indicates that a
metal-insulator transition takes place quite
close to half-filling. 

\section{Isotropic interchain coupling at higher doping}
\label{isodope}

In the following, we consider isotropic interchain coupling
$t_\perp=t$, which is relevant for many ladder materials, 
at a larger doping, $\langle n\rangle=0.875$.
The results are on a $2\times 64$ lattice with $U/t=4$ and at an
inverse temperature $\beta t=5$, which is the lowest temperature we
could reach for these parameters given the size of the average sign.

\subsection{The single-particle spectral weight}
\label{ako_dot}

In Fig.\ \ref{qmc_r0875_64}, we exhibit the single-particle spectral
weight $A({\bf k},\omega)$ for this system.
Two different structures are apparent in the data.
Most of the spectral weight is contained in a cosine-band-like
structure that extends from $\omega\approx -2.8t$ to 
$\omega\approx 1.6t$ in the $k_\perp=0$ channel and from 
$\omega\approx -0.7t$ to $\omega\approx 4.4t$ in the $k_\perp=\pi$
channel.
There is also a broad, incoherent background in both $k_\perp$
channels for photoemission ($\omega<0$) as well as for inverse
photoemission ($\omega>0$) that contains much less spectral weight.
A similar incoherent background has also been found in studies of the
one- and two-dimensional Hubbard model \cite{phl95}.

The dispersive structure which contains the majority of the spectral
weight has the cosine-like form of the noninteracting band structure,
with a bandwidth that agrees well with the $U=0$ bandwidth, shown as a
solid line, for $k_\perp=0$, Fig.\ \ref{qmc_r0875_64}(a), except for
some deviation in the $k_\perp$ channel near $k=\pi$.
In order to try to understand band renormalization effects due to
interactions, one can compare with slave-boson mean-field theory
techniques.\cite{lmh90,zdm96}
However, one finds a narrowing of the noninteracting bands of about
10\%,\cite{ziegler} and a dispersion which does not fit the QMC
results as well as the dispersion of the noninteracting bands.
For higher energies, especially near $k=\pi$ in the $k_\perp=\pi$
channel, the agreement between the QMC data and the noninteracting
dispersion becomes worse.
One can understand this discrepancy in terms of a 
{\it spin-wave-shake-off} effect, which occurs analogously to the
half-filled case.
At half-filling, a self-consistent diagrammatic calculation of the
self-energy within a SDW state leads to additional contributions to the
spectral weight, especially in the center of the magnetic Brillouin
zone, i.e.\ at $k=0$ and $k=\pi$ in both $k_\perp$ channels.
These additional contributions are due to diagram classes involving
transverse magnetic fluctuations. \cite{abk95}
For the doped system, there are still short-range antiferromagnetic
spin correlations which are reminiscent of spin-density-waves,
leading to a spin-wave-shake-off effect in the doped system
near half-filling.

\subsection{Charge and Spin Response}
\label{RPA}

Here we compare the spin and charge excitations of the doped isotropic
system to calculations within the Random Phase Approximation (RPA).
This will allow us to gain some qualitative understanding of many of the 
features of the charge response especially.
The basic idea of the RPA is that the bare Coulomb interaction 
$V^0_{\sigma,\sigma^\prime}({\bf q},\omega_n)$ is replaced by an
effective interaction 
$V^{\rm RPA}_{\sigma,\sigma^\prime}({\bf q},\omega_n)$ in which
screening processes present in the strongly correlated many-body system
are taken into account. 
Analytically summing the ladder diagrams one obtains for the spin and charge
response function:
\begin{equation}
\chi_{c,s}^{\rm RPA}({\bf q},\omega)=
         {\chi_0({\bf q},\omega)\over 1\pm U\chi_0({\bf q},\omega)},
          \label{rpa}
\end{equation}
with
\begin{equation}
\chi_0({\bf q},\omega_n) =
-{1\over N\beta}\sum_{{\bf k},\nu_n}
{\bf G}_{0,{\bf k}}^\sigma(\nu_n-\omega_n)
{\bf G}_{0,{\bf k}+{\bf q}}^\sigma(\nu_n).
\end{equation}
We show the RPA result for the charge susceptibility, calculated on a
$2\times 64$ lattice with $U/t=4$, $\langle n\rangle = 0.875$, and
isotropic $t_\perp=t$ in Fig.\ \ref{Charge_r0875}.
Parts (a) and (c) show the result of the RPA calculation for 
$T\rightarrow 0$ for both values of $q_\perp$, where the poles of 
$\chi_c^{\rm RPA}({\bf q},\omega)$, i.e. the zeros of the denominator
in Eq.\ (\ref{rpa}) are indicated by diamond symbols ($\Diamond$) and
the magnitude of the imaginary part of $\chi_c^{\rm RPA}({\bf q},\omega)$
by the density of the shading.
Figs.\ \ref{Charge_r0875} (b) and (d) show the results of the
QMC/MaxEnt calculation at $\beta t=5$.
The qualitative agreement between the RPA calculation and the
QMC/MaxEnt results 
is quite good, to within the noise caused by the finite resolution and
finite temperature of the QMC/MaxEnt method.
In the $q_\perp=0$ branch, there is a well-defined mode that extends from 
$\omega=0$ at $q=0$ to $\omega\approx 5t$ at $q=\pi$.
By taking the slope of the maximum weighted portion of this mode as 
$q\rightarrow 0$, we obtain a charge velocity $v_c/t=1.74\pm 0.17$.
The band-like structures in the RPA calculation, which contain most
of the spectral weight, clearly have a
similar form to those in the QMC/MaxEnt data.
In addition, there is also a broad spectrum below the well-defined
bands, including gapless excitations at wave vectors 
$q_1=17\pi/32$ and $q_2=25\pi/32$, which are also present in the
QMC/MaxEnt data, although not so well-defined due to the finite
resolution of the QMC/MaxEnt and the relatively small amount of
spectral weight in these modes.

%

The $q_\perp=\pi$ spectra, shown in Figs.\ \ref{Charge_r0875} (c) and
(d), are also quite similar for the RPA and QMC/MaxEnt calculations.
In this channel, most of the spectral weight for both methods is
contained in a well-defined band-like structure which extends from 
$\omega\approx 2.5t$ at $q=0$ to $\omega\approx 6t$ at $q=\pi$.
For the RPA calculation, there is a broad spectrum of excitations below
this band with gapless excitations at $q_3=11\pi/32$ and $q_4=28\pi/32$.
The QMC data, Fig.\ \ref{Charge_r0875}(d), again show very similar
structures.

One can understand the source of the gapless excitations at 
$q_1$ and $q_2$ in the $q_\perp=0$ channel and $q_3$ and $q_4$ in the 
$q_\perp=\pi$ channel in terms of the noninteracting band structure of
the system.
In Fig.\ \ref{band_r0875}, we show the bonding ($k_\perp=0$) and the
antibonding ($k_\perp=\pi$) bands defined by $\varepsilon(k)$ for
$U=0$ with the Fermi wave vectors $k_{F_1}$ and $k_{F_2}$ within each
band indicated.
Two gapless intraband and two gapless interband charge excitations are
possible given this band structure.
The intraband excitations ($q_\perp=0$) occur for $q_2=2\pi-2k_{F_1}$
within the $k_\perp=0$ band and for $q_1=2k_{F_2}$ within the $k=\pi$
band (wrapping back into the reduced Brillouin zone).
The interband transitions ($q_\perp=\pi$) can occur at 
$q_3=k_{F_1}-k_{F_2}$ and $q_4=k_{F_1}+k_{F_2}$.
At these four wave-vectors, gapless particle-particle
excitations can occur, leading to $\omega=0$ poles in 
$\chi_c^{\rm RPA}({\bf q},\omega)$ and thus the gapless excitations
seen in Fig.\ \ref{Charge_r0875} (a) and (c).

In Fig.\ \ref{Spin_r0875}, we show the QMC results as well as the RPA
calculation for the spin susceptibility $\chi_s({\bf q},\omega)$ for
the same parameter values as the charge susceptibility discussed
above, with the results plotted in a similar manner.
While some of the features of the RPA calculation are present, here
the qualitative agreement with the QMC results is not as good.
There is a single gapless ($\omega\rightarrow 0$) mode present in
$q_\perp=0$ channel in the QMC data that increases with increasing
$q$, reaching $\omega\approx 1.5t$ at $q=\pi$.
The RPA calculation, in contrast, shows a broad spectrum which extends
to $\omega\approx 4t$ at $q=\pi$ with the largest spectral weight
contained in two dispersive bands
which become gapless at $q_5\approx 13\pi/32$ and 
$q_6\approx 27\pi/32$.
The deviation of the RPA results from the QMC calculation is probably
due to the following: first, the relatively high temperature of the
QMC simulation ($\beta t=5$) leads to a thermal broadening of the
structures so that individual poles could no longer be resolved, even
with exact (i.e.\ with zero statistical error) data.
Secondly, it is known from DMRG calculations \cite{nws94} that there
is a spin gap in the isotropic system even for moderate doping.
Since the RPA calculation results in a Fermi liquid state, there is no
spin gap in the RPA results, whereas in the QMC data, the presence of
a spin gap, expected from the DMRG results, could lead to the
vanishing of spectral weight at $q_5$ and $q_6$.

In the $q_\perp=\pi$ channel, shown in Fig.\ \ref{Spin_r0875} (c) and
(d), the RPA calculation shows gapless excitations at $q_7\approx
11\pi/32$, $q_8\approx 20\pi/32$ and $q_9\approx 28\pi/32$.
In the QMC data, one can only clearly see the upper part of the
dispersive band that becomes gapless at $q_7$ and has a value
$\omega\approx t$ for $q\rightarrow 0$.
Most of the spectral weight is, in contrast to the RPA calculation,
contained in the $q_\perp=\pi$ channel and is concentrated at  
$\omega\approx 0.5t$ between $q=3\pi/4$ and $q=\pi$.
The overall dispersion of this structure is somewhat cosine--like,
reminiscent of that seen in
Fig.\ \ref{CS_r096875}(b) for small doping and $t_\perp/t=2$ and thus
could be be partially due to the propagation along the ladder of a 
rung spin-triplet excitations in a spin-singlet background, as described
previously in the LRA picture.
Therefore, the QMC data possibly show some hints of a spin gap in
$\chi_s({\bf q},\omega)$ in both $q_\perp$ channels.



\section{CONCLUSION}
\label{CON}

We have discussed the dynamic single-particle, as well as
two-particle (i.e.\ charge and spin) response of a doped two-chain Hubbard
ladder, calculated
using QMC and the Maximum Entropy method and concentrating
on the regime of small doping.
The general evolution of the single--particle as well as the
charge and spin response has been investigated for a filling of 
$\langle n\rangle =0.94$, corresponding to two holes on a $2\times 16$
lattice at $U/t=4.0$, an interaction strength equal to the
one-dimensional bandwidth.
The behavior of the single--particle spectral weight 
$A({\bf k},\omega)$ is dominated by a one--band to two--band
transition as $t_\perp/t$ is decreased from the highly anisotropic
case. 
There are additional interaction effects such as a flattening of the
dispersion near the Fermi level, band narrowing, and reflected bands
due to antiferromagnetic correlations which are especially prominent
near the band transition.
The presence of a single-particle gap when both the bonding and the
antibonding bands cross the Fermi surface, while not resolvable within
the QMC/MaxEnt data, was confirmed using DMRG
calculations.

At large $t_\perp/t$ the dynamical spin and charge susceptibilities
show heavily weighted high--energy features corresponding to local
magnetic and charge excitations on a rung, as well as more lightly
weighted gapless modes corresponding to Luttinger liquid behavior in
the bonding channel.
The presence of the gapless spin mode is also found in the static
magnetic structure factor calculated with the DMRG.
As $t_\perp/t$ is lowered, the spin response evolves first to reflect
the strong short-range antiferromagnetic correlations at 
${\bf q}=(\pi,\pi)$,  and then evolves to the form of uncoupled
one-dimensional chains with similar dispersion and weight in both
transverse channels for small $t_\perp/t$.
The charge response is consistent with a gapless mode in the
$q_\perp=0$ channel as $t_\perp/t$ is reduced.
In the $q_\perp=\pi$ channel, the charge response evolves 
from a form marked by a high-energy local rung excitation to a form
with a low-energy mode with linear dispersion with a weight
distribution similar to that of the $q_\perp=0$ channel as $t_\perp/t$
is reduced.

For the case of one hole at strong interchain coupling, 
$t_\perp \gtsim 2t$ and $U/t=8$, the results are well--described by 
a Local Rung Approximation (LRA), in which the states of one rung are
treated exactly and the interaction between the rungs is treated
perturbatively.
Both the dispersion and the weight distribution in the effective
band structure, given by the single-particle spectral 
weight $A({\bf k},\omega)$, is quantitatively well-fit by a LRA
calculation in which one or two rung singlets are replaced by states
with one hole, and then delocalized into a Bloch band.
The $q_\perp=\pi$ dynamic charge and spin susceptibilities are also
well-described by an LRA picture in which local rung charge or
triplet excitations are delocalized into a band.
However, there is a well-defined gapless charge mode in the
$q_\perp=0$ channel of the charge response signifying metallic
behavior, while spectral weight for $q_\perp=0$ is forbidden by
selection rules in the LRA picture.

For the isotropic system at moderate doping, 
$\langle n\rangle=0.875$, we still find the overall quasi-particle
dispersion essentially unchanged. However, a broad incoherent
background with a width of several $t$, that is analogous to similar
features observed in the one- and two- dimensional Hubbard models, is
found. 
The dispersion of the quasi-particle-like structure
agrees well with that of the $U=0$ band structure.
The general features of the charge susceptibility, including the
origin of wave vectors at which gapless excitations occur, can be
understood in terms of a Random Phase Approximation (RPA) picture.
In contrast, the spin susceptibility shows the signature of short--range
antiferromagnetic ordering not present within the RPA picture.

\section*{ACKNOWLEDGMENTS}

Intensive discussions with D.J.\ Scalapino and R.\ Preuss are
acknowledged and we thank S.\ Haas, H.\ Tsunetsugu and T.M.\ Rice 
for helpful conversations. 
H.E.\ and W.H.\ are grateful to the Bavarian ``FORSUPRA'' program
on high $T_c$ research, to the DFG under Grant No.\ Ha 1537/12-1
and the BMBF Grant No.\ 05 605WWA 6 for financial
support. 
M.\ G.\ Z.\ acknowledges support from the DFN project Tk598-VA/D3, and
R.M.N. was supported by the Swiss National Foundation under
Grant No. 20-46918.96.
The calculations were performed at the HLRZ in
J\"ulich and at the LRZ M\"unchen.

\end{multicols}
\widetext 

\begin{multicols}{2}

\end{multicols}
\newpage
\section*{Figure captions}
\begin{figure}
\caption{
The $U=0$ phase diagram in the $t_\perp$--$\langle n\rangle$--plane.
In the shaded region, both the bonding and antibonding bands intersect
the Fermi level, while in the unshaded region only the bonding band is
occupied.
On the dashed line, the bonding band is half--filled.
The solid points signify parameter values at which we calculate the
dynamical properties in this work.
The open circles represent points at which we calculate the
quasiparticle gap with the DMRG.
}
\label{U0phase}
\end{figure}
\begin{figure}
\caption{
 The single-particle spectral weight $A({\bf k},\omega)$ for
$t_\perp/t=2.2$, $U/t=4$ and $\beta t=8$ on a $2\times 16$ lattice at a
filling of $\langle n\rangle=0.94$. 
The two $k_\perp$ branches are overlayed on the same plot since they
are well-separated in energy, with the open and filled symbols marking
the position of the $k_\perp=0$ and $k_\perp=\pi$ peaks, respectively,
and the error bars indicating the width of the peaks.
The density of shading indicates the amount of the spectral weight.
}
\label{Akwtp2p2}
\end{figure}
\begin{figure}
\caption{
$A({\bf k},\omega)$ plotted as in Fig.\ 2, with the same parameters,
except with $t_\perp/t=1.6$.
}
\label{Akwtp1p6}
\end{figure}
\begin{figure}
\caption{
$A({\bf k},\omega)$ plotted as in Figs.\ 2-3, 
except with $t_\perp/t=1.0$. 
}
\label{Akwtp1}
\end{figure}
\begin{figure}
\caption{
$A({\bf k},\omega)$ plotted as in Figs.\ 2-4, 
except with $t_\perp/t=0.2$ and (a) $k_\perp=0$ and (b) $k_\perp=\pi$
plotted separately.
}
\label{Akwtpp2}
\end{figure}
\begin{figure}
\caption{
Finite size extrapolation of the quasiparticle gap, calculated using
the DMRG for $U=4$, $\langle n \rangle=0.9375$, and $t_\perp/t=1.3,1.6$.
}
\label{quasigap}
\end{figure}
\widetext

\begin{figure}
\caption{
(a) and (b), the spin susceptibility $\chi_s({\bf q},\omega)$ and (c)
and (d), the charge
susceptibility $\chi_c({\bf q},\omega)$ for $t_\perp/t=2.2$, $U/t=4$
and $\beta t=8$ on a $2\times 16$ lattice at a 
filling of $\langle n\rangle=0.94$.
}
\label{Chitp2p2}
\end{figure}

\widetext

\begin{figure}
\caption{
The spin and charge susceptibilities plotted as in Fig.\ 7, except
for $t_\perp/t=1.6$.
}
\label{Chitp1p6}
\end{figure}
\widetext

\begin{figure}
\caption{
The spin and charge susceptibilities plotted as in Figs.\ 7-8, except
for $t_\perp/t=1.0$.
}
\label{Chitp1}
\end{figure}
\widetext

\begin{figure}
\caption{
The spin and charge susceptibilities plotted as in Fig.\ 7-9, except
for $t_\perp/t=0.2$.
The straight lines indicate the spin- and charge velocities of the
corresponding 1D Hubbard chain (from Ref.\ 46). 
}
\label{Chitpp2}
\end{figure}
\begin{figure}
\caption{
The spin structure factor $S(q)$ for $t_\perp/t=1.6,2.2$ calculated with the
DMRG with $U/t=4$ and $\langle n \rangle = 0.9375$.
}
\label{Sqn94}
\end{figure}
\begin{figure}
\caption{
The single-particle spectral weight $A({\bf k},\omega)$ for
$t_\perp/t=2$, $U/t=8$ and $\beta t=8$ on a $2\times 16$ lattice at a
filling of $\langle n\rangle=0.96875$ (1 hole). 
The error bars denote the width of the peak obtained using the Maximum
Entropy technique.
The dotted lines indicate the dispersion of the noninteracting, $U=0$
bands, and the solid line in (a) indicates the dispersion obtained
from the LRA calculation.
}
\label{xyz}
\end{figure}
\begin{figure}
\caption{The spin and charge susceptibilities ($\chi_s({\bf q},\omega)$: parts
(a) and (b), $\chi_c({\bf q},\omega)$: parts (c) and (d)) for
$t_\perp/t=2$, $U/t=8$ and $\beta t=8$ on a $2\times 16$ lattice at a
filling of $\langle n\rangle=0.96875$ (1 hole).
The $q_\perp=0$ channel is shown in (a) and (c), while the
$q_\perp=\pi$ channel is shown in (b) and (d).
}
\label{CS_r096875}
\end{figure}
\begin{figure}
\caption{The single-particle spectral weight $A({\bf k},\omega)$ for
$t_\perp/t=1$, $U/t=4$ and $\beta t=5$ on a $2\times 64$ lattice at a
filling of $\langle n\rangle=0.8776\pm 0.0001$.
The $k_\perp=0$ and $k_\perp=\pi$ data are shown 
in parts (a) and (b), respectively. 
The solid line represents the dispersion of the noninteracting ($U=0$)
bands.
}
\label{qmc_r0875_64}
\end{figure}
\begin{figure}
\caption{
The charge susceptibility $\chi_c({\bf q},\omega)$ for isotropic
$t_\perp=t$ with $U/t=4$ on a $2\times 64$ lattice at a filling of
$\langle n\rangle=0.875$.
Parts (a) and (c) show the results of the RPA calculation for
both $q_\perp$ channels at $T=0$ (grey-shaded data and diamond
$\Diamond$ symbols), while parts (b) and (d) show the QMC results for
$\beta t=5$.}
\label{Charge_r0875}
\end{figure}
\begin{figure}
\caption{
Band structure of the interaction free case ($U=0$) of a $2\times 64$
system with $t_\perp/t=1.0$ and a particle density of $\langle
n\rangle=0.875$.}
\label{band_r0875}
\end{figure}
\begin{figure}
\caption{
The spin susceptibility  $\chi_s({\bf q},\omega)$ for isotropic
$t_\perp=t$ with $U/t=4$ on a $2\times 64$ lattice at a filling of
$\langle n\rangle=0.875$.
Parts (a) and (c) show the results of the RPA calculation for
both $q_\perp$ channels at $T=0$ (grey-shaded data and diamond
$\Diamond$ symbols), while parts (b) and (d) show the QMC results for
$\beta t=5$.}
\label{Spin_r0875}
\end{figure}

\end{document}